\documentclass[conference]{IEEEtran}

\ifCLASSINFOpdf
  \usepackage[pdftex]{graphicx}
  \DeclareGraphicsExtensions{.pdf,.jpeg,.png}
\else
\fi

\usepackage{amsmath}
\usepackage{array}
\ifCLASSOPTIONcompsoc
  \usepackage[caption=false,font=normalsize,labelfont=sf,textfont=sf]{subfig}
\else
  \usepackage[caption=false,font=footnotesize]{subfig}
\fi

\usepackage[hyphens]{url}
\usepackage{hyperref}

\usepackage{xcolor}
\hypersetup{
    colorlinks,
    linkcolor={red!50!black},
    citecolor={blue!50!black},
    urlcolor={blue!80!black}
}

\usepackage{booktabs}
\usepackage{varioref}
\usepackage{cleveref}
\usepackage{color}

\usepackage{listings}
\usepackage[utf8]{inputenc}

\definecolor{codegreen}{rgb}{0,0.6,0}
\definecolor{codegray}{rgb}{0.5,0.5,0.5}
\definecolor{codepurple}{rgb}{0.58,0,0.82}
\definecolor{backcolour}{rgb}{0.95,0.95,0.92}

\lstdefinestyle{mystyle}{
    backgroundcolor=\color{backcolour},   
    keywordstyle=\color{codegreen},
    numberstyle=\tiny\color{codegray},
    stringstyle=\color{codepurple},
    basicstyle=\tiny,
    breakatwhitespace=false,         
    breaklines=true,                 
    captionpos=b,                    
    keepspaces=true,                 
    numbersep=5pt,                  
    showspaces=false,                
    showstringspaces=false,
    showtabs=false,                  
    tabsize=2
}
\usepackage[numbers,sort]{natbib}

\usepackage{balance}

\begin{document}

\title{Scheduling With Inexact Job Sizes:\\The Merits of Shortest Processing Time First}

\author{\IEEEauthorblockN{Matteo Dell'Amico}
\IEEEauthorblockA{Symantec Research Labs}}

\maketitle

\begin{abstract}
It is well known that \emph{size-based} scheduling policies, which take into account job size (i.e., the time it takes to run them), can perform very desirably in terms of both response time and fairness.
Unfortunately, the requirement of knowing \textit{a priori} the exact job size is a major obstacle which is frequently insurmountable in practice.
Often, it is possible to get a coarse \emph{estimation} of job size, but unfortunately analytical results with inexact job sizes are challenging to obtain, and simulation-based studies show that several size-based algorithm are severely impacted by job estimation errors.
For example, Shortest Remaining Processing Time (SRPT), which yields optimal mean sojourn time when job sizes are known exactly, can drastically underperform when it is fed inexact job sizes.

Some algorithms have been proposed to better handle size estimation errors, but they are somewhat complex and this makes their analysis challenging. We consider Shortest Processing Time (SPT), a \emph{simplification} of SRPT that skips the update of ``remaining'' job size and results in a preemptive algorithm that simply schedules the job with the shortest estimated processing time. When job size is inexact, SPT performs comparably to the best known algorithms in the presence of errors, while being definitely simpler. In this work, SPT is evaluated through simulation, showing near-optimal performance in many cases, with the hope that its simplicity can open the way to analytical evaluation even when inexact inputs are considered.\end{abstract}

\IEEEpeerreviewmaketitle

\section{Introduction}

Scheduling is a pervasive problem in computer science, related to the handling of queues for accessing a shared resource (e.g., processor time or network bandwidth); \emph{size-based} scheduling techniques are those that use information about \emph{job size} (i.e., how long the job will keep the shared resource busy) to take scheduling decisions.

Size-based scheduling is known to perform very well: \citet{schrage1966queue} proved in \citeyear{schrage1966queue} that Shortest Remaining Processing Time (SRPT), a size-based policy which prioritizes jobs closest to completion, minimizes the mean sojourn time. Further works \cite{Friedman2003,dell2016psbs} have shown a class of size-based algorithms that guarantee strong fairness properties while performing near-optimally.

As discussed in Section~\ref{sec:related}, even though these properties are very appealing, size-based scheduling is not widely used in practice: knowing exactly the job size in advance is often impossible or impractical, and performance may degrade noticeably when scheduling is based on estimated job size \cite{lu2004size}. Some algorithms, such as PSBS (Practical Size Based Scheduling)~\cite{dell2016psbs} and MCSS (Modified Comparison Splitting Scheduling)~\cite{mailach2017scheduling}, have been shown to perform well even if job size estimation errors are present. Unfortunately, these algorithms are non-trivial, and system designers might be rightfully skeptical of including them in their systems. Additionally, the problem of size-based scheduling in presence of job size estimation errors appears difficult to handle analytically, and nontrivial scheduling algorithms do not ease analysis.

In this work, PSBS and MCSS are taken into account together with other non size-based scheduling algorithms described in Section~\ref{sec:algorithms}, and compared with a very simple algorithm: Shortest Processing Time (SPT), which simply prioritizes the job with the smallest (estimated) job size. SPT is actually a simplification of SRPT, because it doesn't update job (remaining) sizes when work is performed; it can be implemented with a simple priority queue data structure. The above algorithms are evaluated using Schedsim~\cite{dell2013simulator}; experimental settings are discussed in Section~\ref{sec:settings}.

The experimental results of Section~\ref{sec:results} confirm the previous findings on PSBS~\cite{dell2016psbs} and MCSS~\cite{mailach2017scheduling} and allow to compare the two algorithms, showing that PSBS generally outperforms MCSS. Perhaps surprisingly, SPT performs excellently, being essentially equivalent to PSBS in terms of both response times and fairness.

Section~\ref{sec:discussion} elaborates on the results, to facilitate an intuitive understanding that can justify the experimental results.

Section~\ref{sec:conclusion} concludes by discussing how SPT conjugates simplicity with near-optimal performance in a large set of realistic cases. Hopefully, the results of this work will both stimulate the adoption of SPT or similar algorithms in real systems and enable better analytic treatment of scheduling performance in presence of errors.

\section{Related Work}
\label{sec:related}


While many works have studied scheduling for single-server queues, most of them focused on the extreme cases where job size is either completely unknown (\emph{size-oblivious} algorithms) or known perfectly. When the job size distribution is skewed---meaning that resources are occupied most of the time by a minority of large jobs---algorithms such as Least Attained Service (LAS)~\cite{rai2003analysis} or multi-level queues~\cite{kleinrock1976queueing,guo2002scheduling} can still perform well by prioritizing new jobs. LAS is evaluated experimentally in this work; for more details about it, see Section~\ref{sec:las}.

\citet{lu2004size} were the first to consider estimation errors for size-based scheduling, and observed that existing algorithms perform well only when job sizes were rather accurately estimated. Further work~\cite{dell2014revisiting,dell2016psbs} has shown that most problems happen when the job size distribution is skewed and large jobs' sizes are under-estimated: in that case, these jobs eventually reach a very high priority and are not preempted when smaller jobs arrive, clogging the system. PSBS~\cite{dell2016psbs} (Section~\ref{sec:psbs}) and MCSS~\cite{mailach2017scheduling} (Section~\ref{sec:cs}) are proposals that perform better on estimated job sizes; in Section~\ref{sec:results} both are evaluated and compared to SPT.

Unfortunately, analytic results on scheduling based on inexact size are few: \citet{wierman2008scheduling} assume bounded errors and give pessimistic performance bounds, and some other works estimate mean sojourn time in running systems when the job size distribution is unknown for the last-come-first-serve~\cite{meykhanadzhyan_new_2016}, processor-sharing~\cite{horvath_estimating_2017} and first-come-first-serve~\cite{milovanova_bounding_2018} schedulers. The relative scarcity of work in this space suggests that this kind of analysis is difficult; hopefully, a simple algorithm like SPT may be a good target for analysis.



In the literature, some systems use job size estimation to drive scheduling. For batch computation systems, a part of jobs is run to estimate running time~\cite{wolf2010flex,pastorelli2013hfsp}; web servers use file size to estimate serving time~\cite{schroeder2006web}. More elaborate approaches predict the size of database queries~\cite{lipton1995query}, MapReduce jobs~\cite{ARIA11,nsdi12-c,query_perf}, deep learning training~\cite{peng2018optimus} and the length of call-center calls~\cite{emadi_can_2019}: approaches such as these can be used to inform size-based schedulers.


\section{Evaluated Algorithms}
\label{sec:algorithms}

We evaluate MCSS and SPT against a few size-oblivious and size-based scheduling algorithms chosen between the best performers in~\cite{dell2016psbs}. All these algorithms are \emph{preemptive}, i.e., resources can be reclaimed from currently running algorithms; in skewed workloads, non-preemptive algorithms perform definitely worse because small jobs cannot interrupt large running ones. 

\subsection{Size-Oblivious Algorithms}

Two size-oblivious algorithms are considered in this work: \emph{Processor Sharing} (PS) and Least Attained Service (LAS).

\subsubsection{PS}
the most common algorithm in many concrete scheduler implementations; it is often considered fair, because each running job is assigned an equal share of resources in the system. The other side of the coin is that, when the system is loaded, every job progresses slowly.

\subsubsection{LAS~\cite{rai2003analysis}}
\label{sec:las}
also known as Fore\-ground-Back\-ground~\cite{kleinrock1975theory} and Shortest Elapsed Time~\cite{coffman1973operating},
prioritizes the job(s) that has been served the least resources; if several jobs are tied, they are served an equal share of resources each. LAS performs well in the common case when the job size distribution is skewed: most jobs are small and can get served quickly if they get access to resources; by preempting running jobs aggressively, LAS generally ensures a short response time for small jobs. Analytical results~\cite{yashkov_processor-sharing_1987,righter_scheduling_1989} prove that LAS has optimal mean sojourn time among size-oblivious strategies for several heavy-tailed job size distributions.
\subsection{Size-Based Algorithms}

\subsubsection{SRPT}
\label{sec:srpt}
this is the scheduling policy that has optimal mean sojourn time when it is fed exact job sizes as input~\cite{schrage1966queue}. Unfortunately, when the inputs are approximated job sizes (in the following, we will refer to this case as SRPTE, where the trailing `E' is used as a reminder that the scheduler is fed input with errors), this algorithm can severely underperform~\cite{lu2004size,dell2016psbs}.

\subsubsection{PSBS~\cite{dell2016psbs}}
\label{sec:psbs}
a generalization of FSPE+PS~\cite{dell2014revisiting}, which in turn generalizes FSP~\cite{Friedman2003}. FSP serves jobs in the order in which they would terminate in a virtual system running PS, used as a reference for fairness. Without estimation errors, FSP is guaranteed to complete each job before PS. FSPE+PS handles job size estimation errors by avoiding that large underestimated jobs starve small ones; PSBS improves on FSPE+PS thanks to a more efficient $O(\log n)$ implementation and by supporting job priorities. Behaviors of PSBS, FSPE+PS and FSP are undistinguishable when fed exact job size information and all jobs belong to the same priority class.

\subsubsection{Comparison Splitting Algorithms}
\label{sec:cs}

Comparison Splitting (CS)~\cite{jelenkovic2007adaptive} is an adaptive algorithm that prioritizes small jobs by assigning them to classes: when a new job is submitted, it is assigned to class $i$, where $i$ is the number of jobs between the last $r$ submitted ones that had smaller size than it. At each time, a job in the smallest non-empty queue is served.

CS approximates SPT; however, when the workload has a skewed size distribution, smaller jobs in the $r$\textsuperscript{th} queue can be starved by the largest ones which remain in the system for a long time~\cite{mailach2017scheduling}. For this reason, \citet{mailach2017scheduling} proposed Modified Comparison Splitting Scheduling (MCCS), in which the $r$\textsuperscript{th} queue is a LIFO queue rather than a FIFO, avoiding this starving phenomenon.

\subsubsection{SPT}
it is, in principle, similar to CS and MCSS: the job with the smallest size is served; the amount of time a job has been served is simply not taken into account. SPT is very easy to implement: a data structure implementing a min-queue (e.g., a binary heap) can be used to take each scheduling decision quickly in $O(\log n)$ time. SPT should not be confounded with Shortest Job First (SJF), which is not preemptive; it is sometimes known in literature as Preemptive Shortest Job First (PSJF)~\cite{wierman_classifying_2003}, but for clarity we avoid that term because it is sometimes used as a synonym for SRPT~\cite{coffman_jr_computer_1968}. SPT can be seen as a preemptive, infinite-class version of the $c\mu$-rule~\cite{cox_queues_1991}.

\section{Experimental Settings}
\label{sec:settings}

Schedsim~\cite{dell2013simulator}, an open-source\footnote{\url{https://github.com/bigfootproject/schedsim}} Python discrete event simulator, is used in this work to evaluate the above algorithms

\begin{table}[!t]
    \centering
    \caption{Default simulation parameters, as in~\cite{dell2016psbs}.}
    \label{tab:params}
    \begin{tabular}{llr}
        \toprule
        Parameter & Description & Default \\
        \midrule
        sigma & $\sigma$ in the log-normal error distribution & 0.5 \\
        shape & shape for Weibull job size distribution & 0.25 \\
        timeshape & shape for Weibull inter-arrival time & 1 \\
        njobs & number of jobs in a workload & 10\ 000 \\
        load & system load & 0.9 \\
        \bottomrule
    \end{tabular}
\end{table}

Schedsim generates synthetic workloads according to a set of parameters; Table~\vref{tab:params} reports the default value of each: when not explicitly mentioned otherwise, those are the values used in the experiments of Section~\ref{sec:results}. The most important parameters are \emph{sigma} and \emph{shape}: they represent, respectively, the level of \emph{precision} of job size estimations, and the \emph{skew} of the job size distribution. We use the same default values used in~\cite{dell2016psbs}, to allow comparing directly the results between this paper and that work. We remark the use of a high load, 0.9, to stress the scheduling algorithms.

Estimation error has a log-normal distribution; this has an intuitive justification, because over- and under-estimation by the same multiplicative factor are equally likely and negative job size estimation is impossible, and an empirical one: \citet{pastorelli2013hfsp} found their empirical error distribution was indeed fitting a log-normal. Sigma expresses a large range of estimation errors: the minimum value, 0.125, reflects a median relative error of 1.088; the maximum value, 4, corresponds to a median relative error of 14.85: job sizes are mis-estimated by more than an order of magnitude.

For synthetic workloads, job size distribution is represented by the shape parameter of a Weibull distribution. Values less than 1 correspond to heavy-tailed distributions (for our minimum value, 0.125, the distribution is extremely skewed); for a value of 1 we obtain an exponential distribution; as this value grows we obtain bell-shaped curves that are more and more narrow. We confirm our experimental results with real workloads as well.

Like \citet{mailach2017scheduling} do, CS and MCSS are configured with $r=10$. Each data point shown in the experiments is obtained by averaging at least 30 individual experiment runs.

\section{Experimental Results}
\label{sec:results}

In this Section, CS, MCSS and SPT are evaluated and compared with SRPT, PSBS, LAS and PS on synthetic and real workloads. Performance is evaluated through \emph{mean sojourn time} (MST), which is the average time between when a job is submitted and when it is completed; for fairness \emph{slowdown} is taken into account: it is the ratio between its sojourn time and the one it would have if not constrained by the scheduled resource. Following the convention of~\cite{dell2016psbs}, SRPTE, CSE, MCSSE, and SPTE are respectively SRPT, CS, MCSS and SPT when fed inaccurate size information. PSBS is considered to use inaccurate size information, as when size information is correct the algorithm is indistinguishable from FSP.

\subsection{Performance}

\begin{figure*}[!htbp]
    \centering
    \subfloat[CSE.]{%
        \includegraphics[width=\columnwidth]{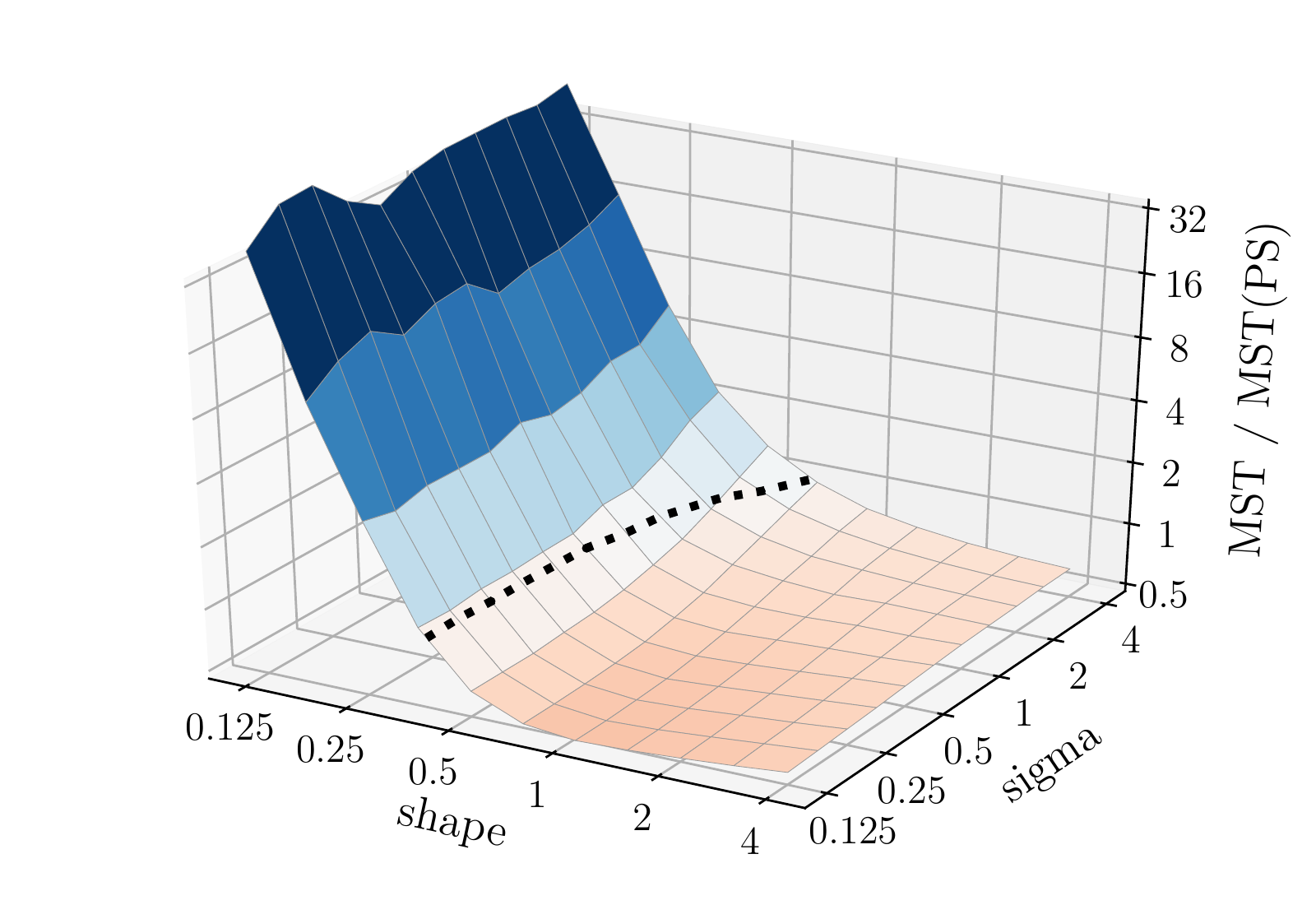}%
        \label{fig:cse_3d}}
    \subfloat[MCSSE.]{%
        \includegraphics[width=\columnwidth]{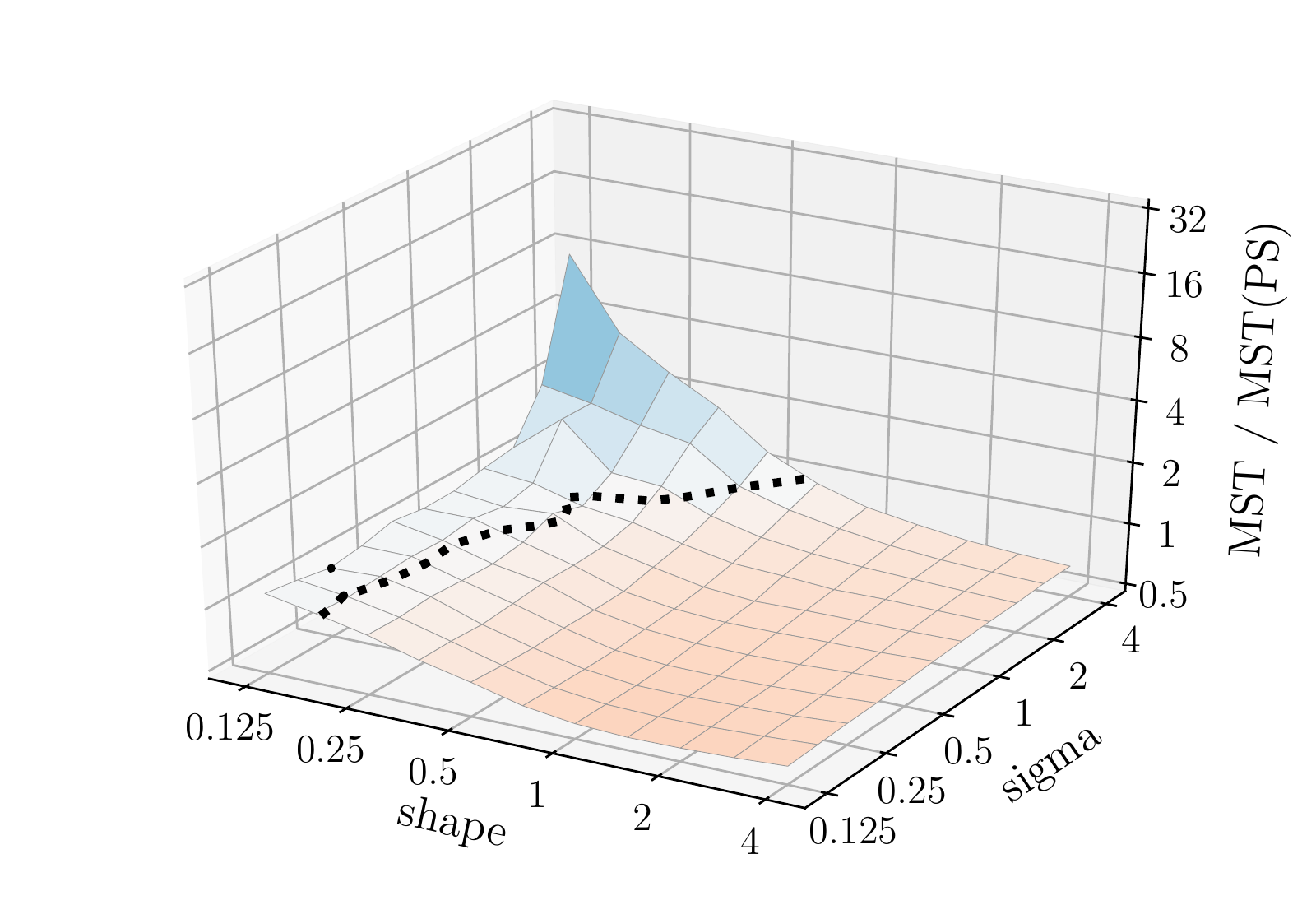}%
        \label{fig:mcsse_3d}}
    \\
    \vspace{-0.45cm}
    \subfloat[PSBS.]{%
        \includegraphics[width=\columnwidth]{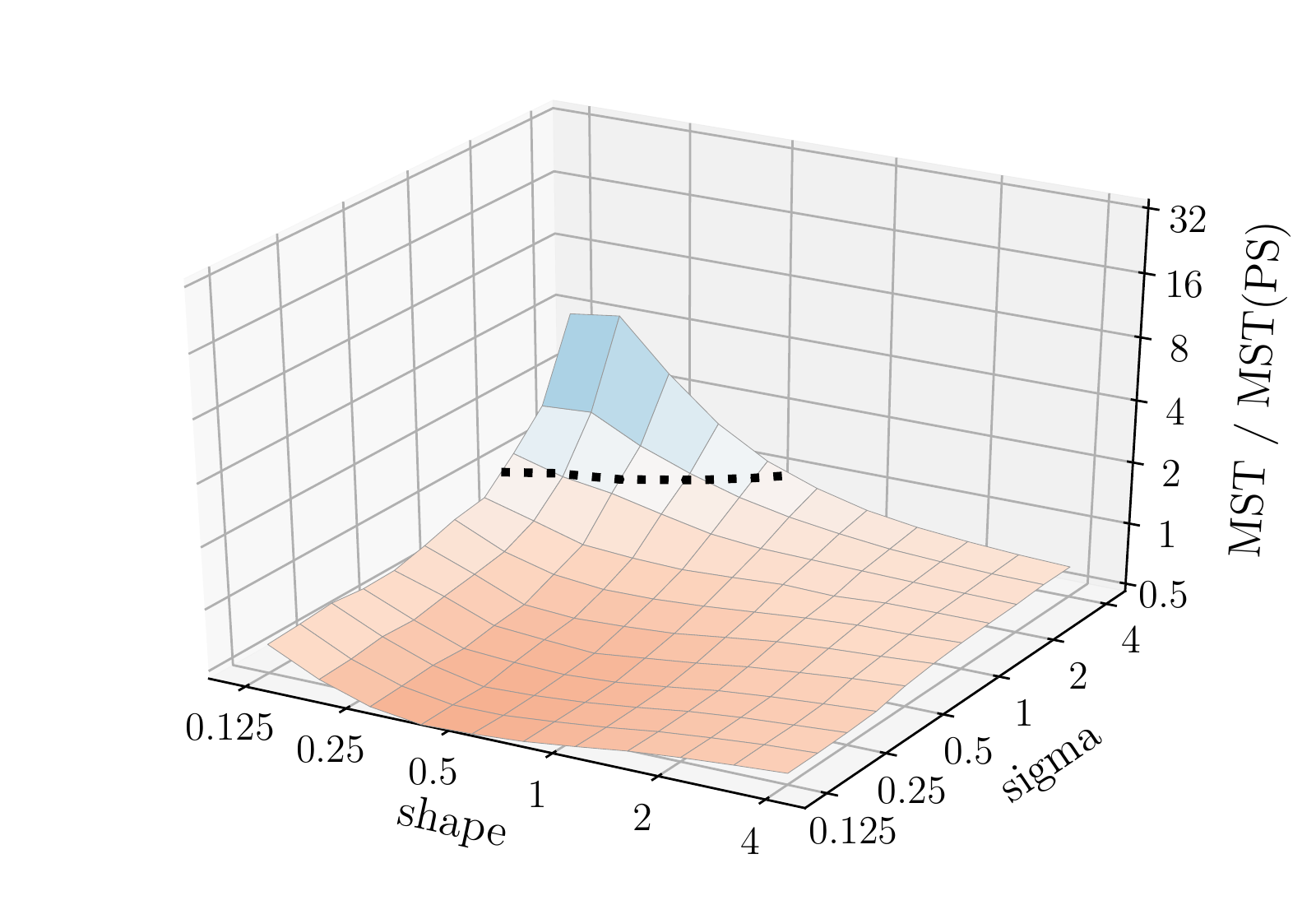}%
        \label{fig:psbs_3d}}
    \subfloat[SPTE.]{%
        \includegraphics[width=\columnwidth]{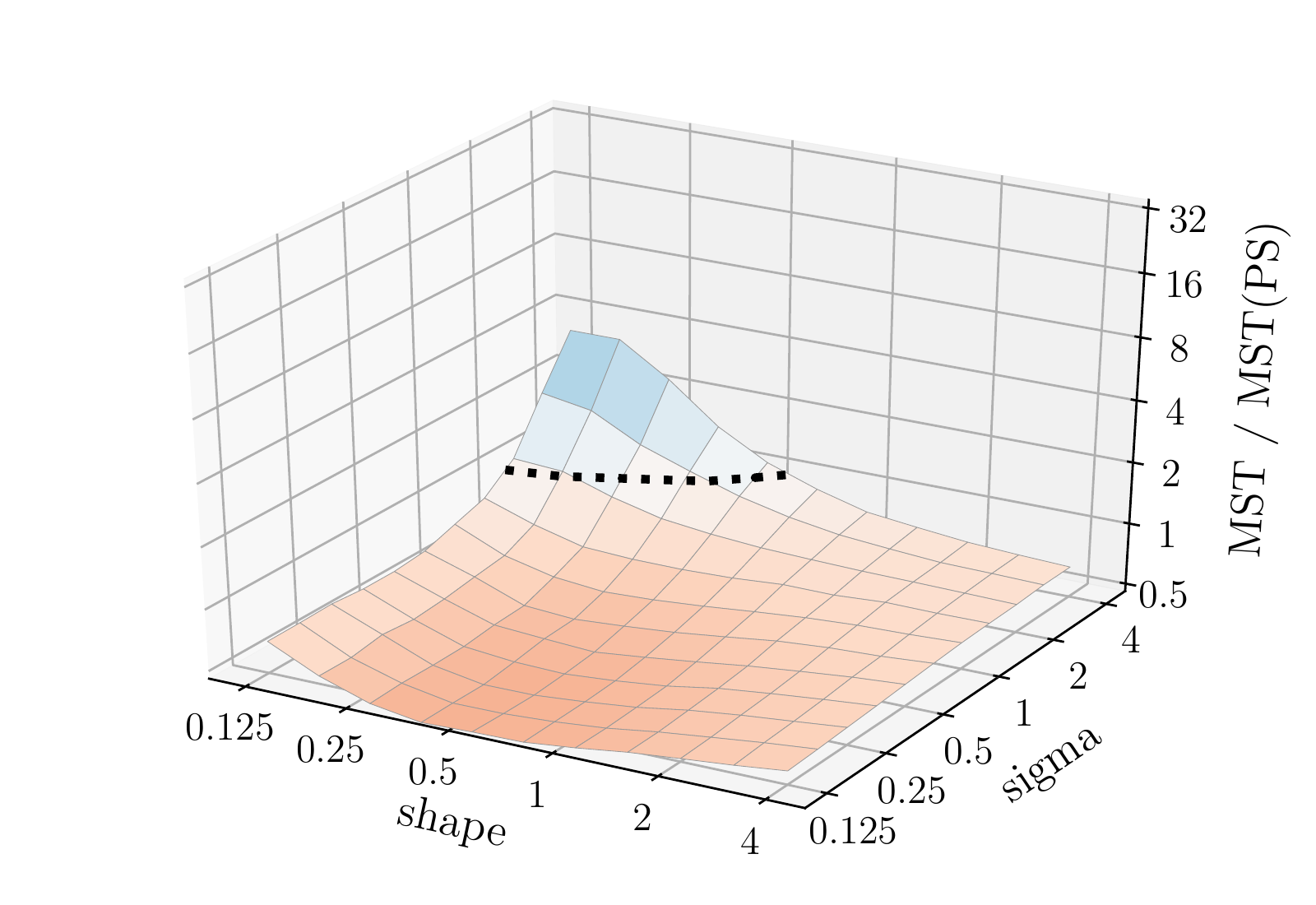}%
        \label{fig:spte_3d}}
    \caption{Mean sojourn time (MST) of various algorithms against PS: at the dotted line, MST is equivalent to PS. A low shape value corresponds to high job size skew, while high sigma implies imprecise size estimates.}
    \label{fig:3d}
\end{figure*}

Our analysis begins by comparing size-based scheduling algorithms to PS, the typical baseline for fair scheduling. In Figure~\vref{fig:3d}, the ratio between the MST of these algorithms and the MST of PS is shown, while varying shape (job size skew) and sigma (size estimation precision). A value lower than 1 means that these algorithms outperform PS; conversely, values higher than 1 means that these algortihms have worse MST than PS. 

The results on CSE and MCSSE confirm those of \citet{mailach2017scheduling}: when the job size distribution is skewed, CSE (Figure~\ref{fig:cse_3d}) performs very badly because the largest jobs starve the system; interestingly, this result applies irrespective of the job size estimation error. The small modification that leads to MCSSE appears quite effective, yielding better performance overall (Figure~\ref{fig:mcsse_3d}); still, when job size estimation errors are not very large, MCSSE performs similarly to PS for very skewed workloads. PSBS and SPTE (Figures~\ref{fig:psbs_3d} and~\ref{fig:spte_3d}) perform best, behaving better than PS in almost all cases except those where the workload is very skewed \emph{and} job size estimation is very imprecise.

\begin{figure}[!htbp]
    \centering
    \includegraphics[width=\columnwidth]{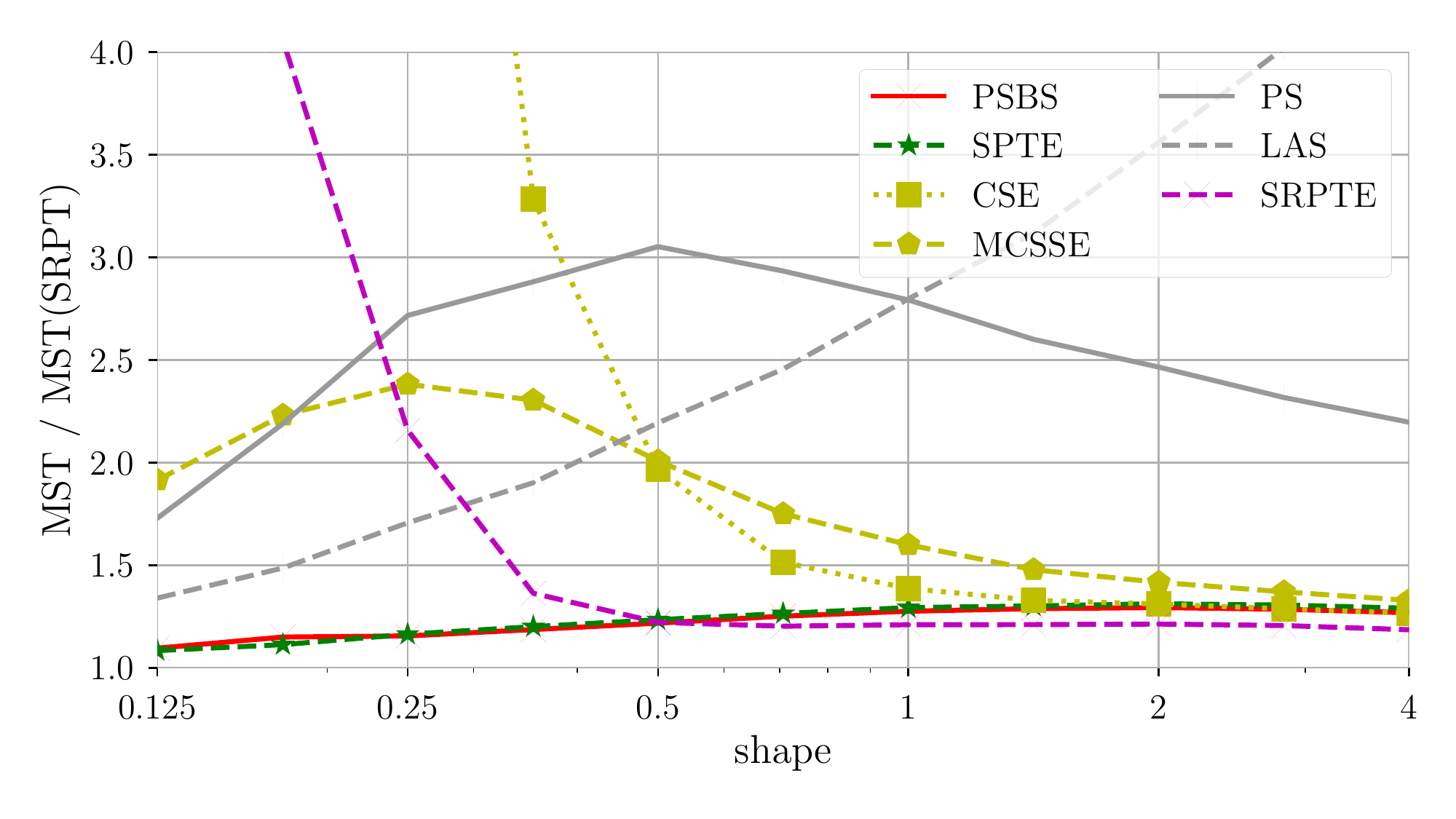}
    \caption{Impact on mean sojourn time of the Weibull distribution's shape parameter. Values lower than 1 corresponds to heavy-tailed workloads, higher than 1 to light-tailed ones, and 1 is the exponential distribution.}
    \label{fig:shape}
\end{figure}

To better understand the impact of job size skew, we now fix sigma at its default value of 0.5 and compare the performance of our algorithms against the optimal one, SRPT, with \emph{exact} job size information. Figure~\vref{fig:shape} shows how these algorithms perform by normalizing MST against the optimal value of SRPT: a value of 1 would be optimal.

We can see how both PSBS and SPTE perform close to optimally, with an MST which is never more than 1.3 times the optimal one of SRPT with \emph{exact} job size information. MCSSE is at worst similar to PS, while we see again how CSE underperforms for highly skewed workloads. The SRPTE policy (i.e., SRPT with inexact job sizes as inputs) has good performance when job sizes are somehow uniform, but severely underperforms when the size distribution is very skewed (shape $< 0.5$).

\begin{figure}[!htbp]
    \centering
    \includegraphics[width=\columnwidth]{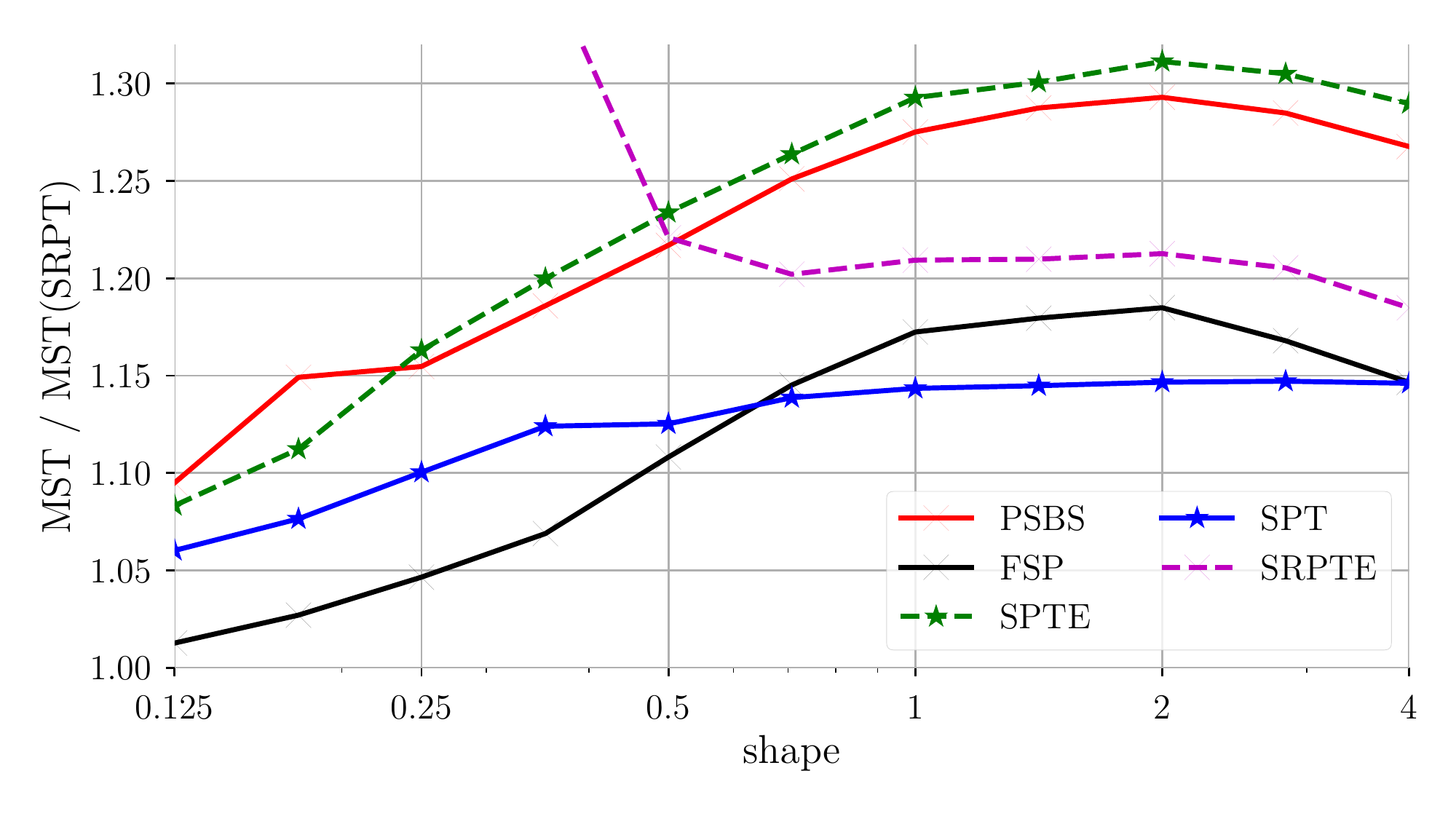}
    \caption{SPTE and PSBS compared to their performance without errors (SPT and FSP respectively).}
    \label{fig:noerror}
\end{figure}

Figure~\vref{fig:noerror} zooms in on the impact of estimation errors for SPTE and PSBS, comparing them to the very same algorithms using correct size information: SPT and FSP, respectively.
%
It is interesting to see how there is some difference between the MST of SPT and FSP (FSP can perform better, up to around 5\%, on skewed workloads; SPT appears preferable by a similar margin on quite homogeneous workloads). An explanation can be that in highly skewed workloads the bulk of the load is due to very few large jobs; the majority of short jobs are served very quickly, queues are short, and aging in FSP reduces job sizes similarly to the optimal SRPT policies. Conversely, when job sizes are more uniform, it is more likely that multiple jobs of similar size are present in the queue; enforcing FSP's ageing may result in serving jobs that are older but larger. In any case, the difference in MST drops to around 1-2\% when estimation errors enter in the picture for PSBS and FSP, suggesting that estimation errors trump these subtle differences in scheduling choices.

\begin{figure}[!htbp]
    \centering
    \subfloat[shape=0.25.]{%
        \includegraphics[width=\columnwidth]{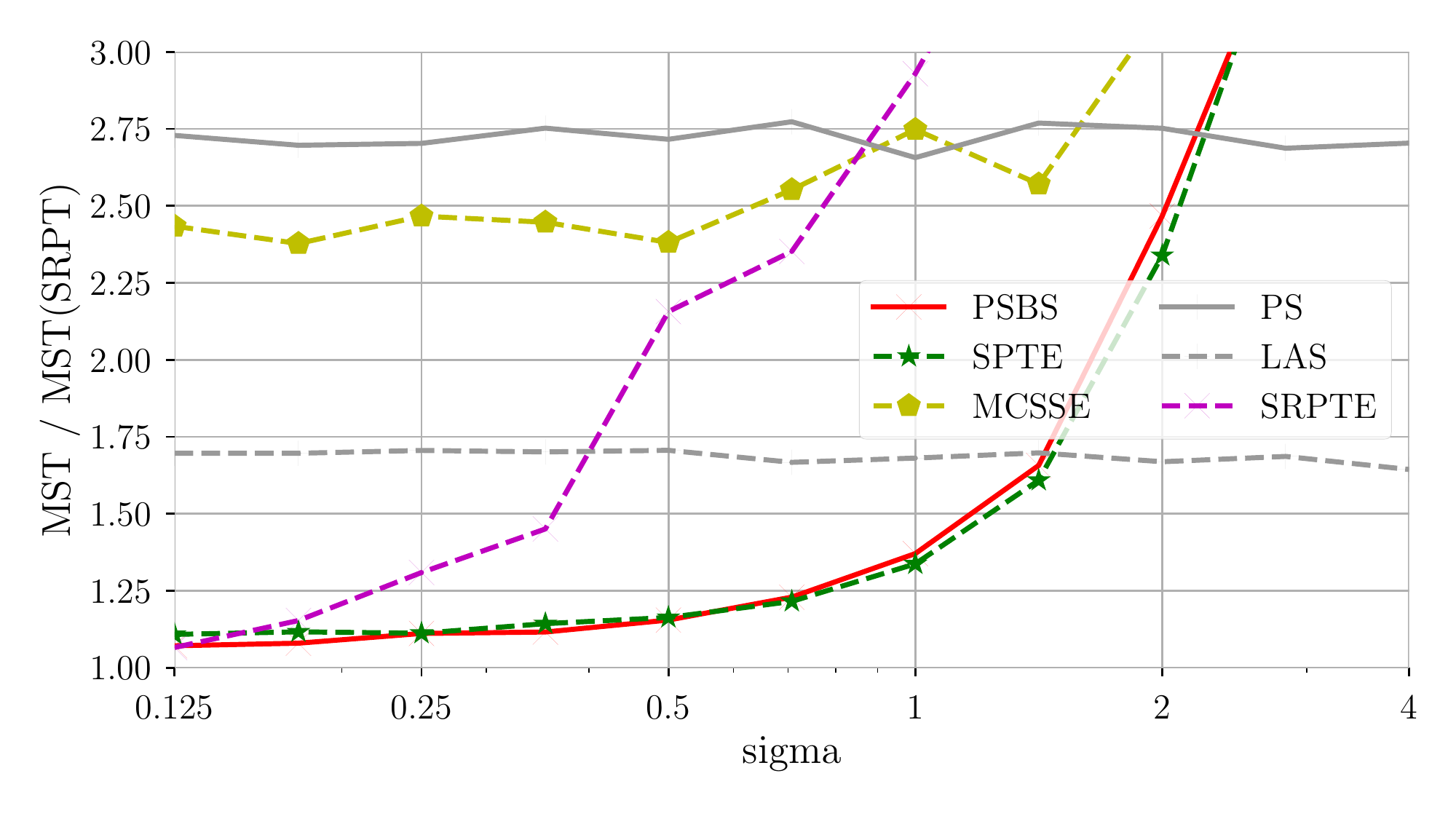}%
        \label{fig:sigma1}}
    \\
    \subfloat[shape=0.177.]{%
        \includegraphics[width=\columnwidth]{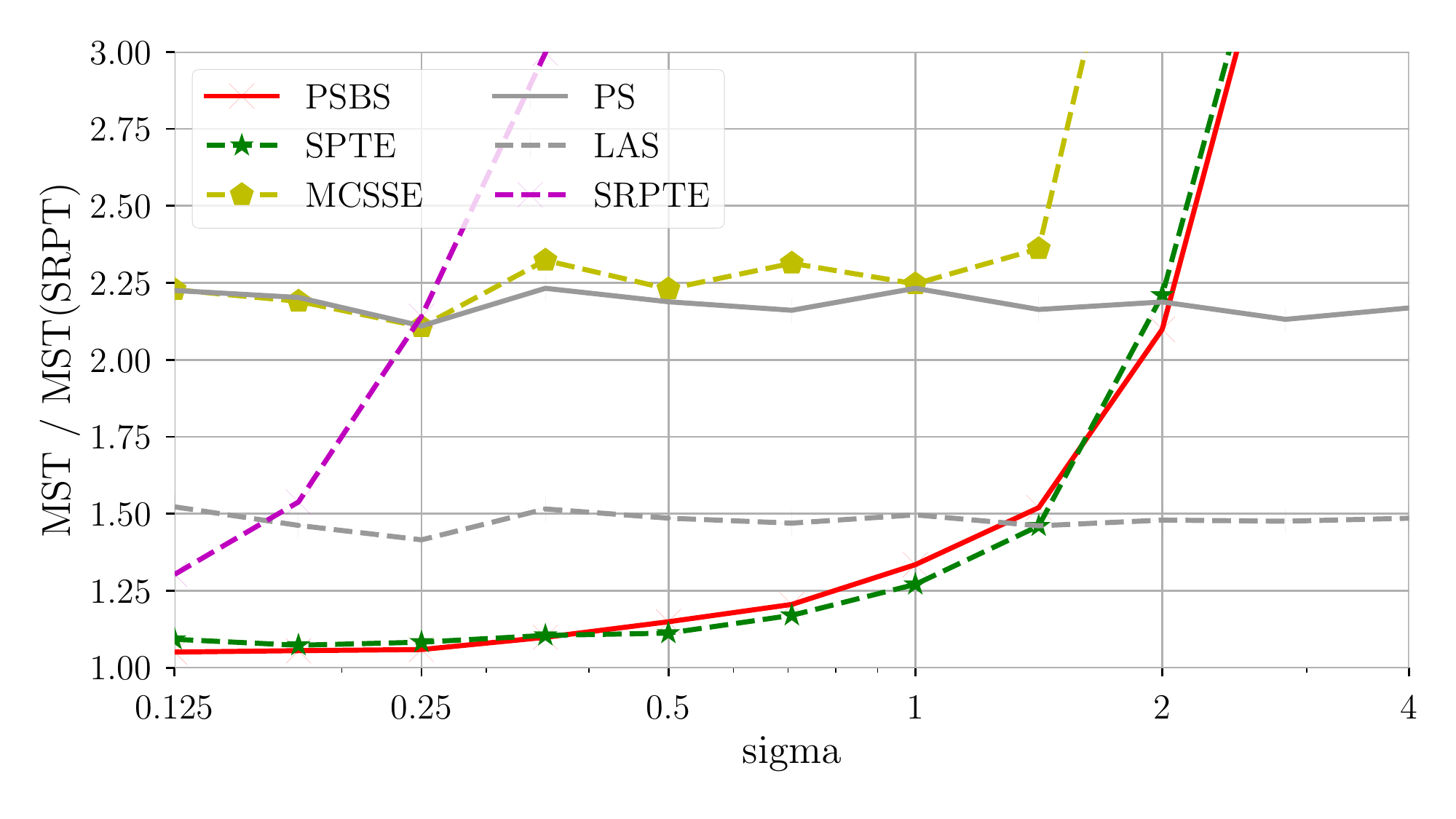}%
        \label{fig:sigma2}}
    \\
    \subfloat[Largest skew: shape=0.125.]{%
        \includegraphics[width=\columnwidth]{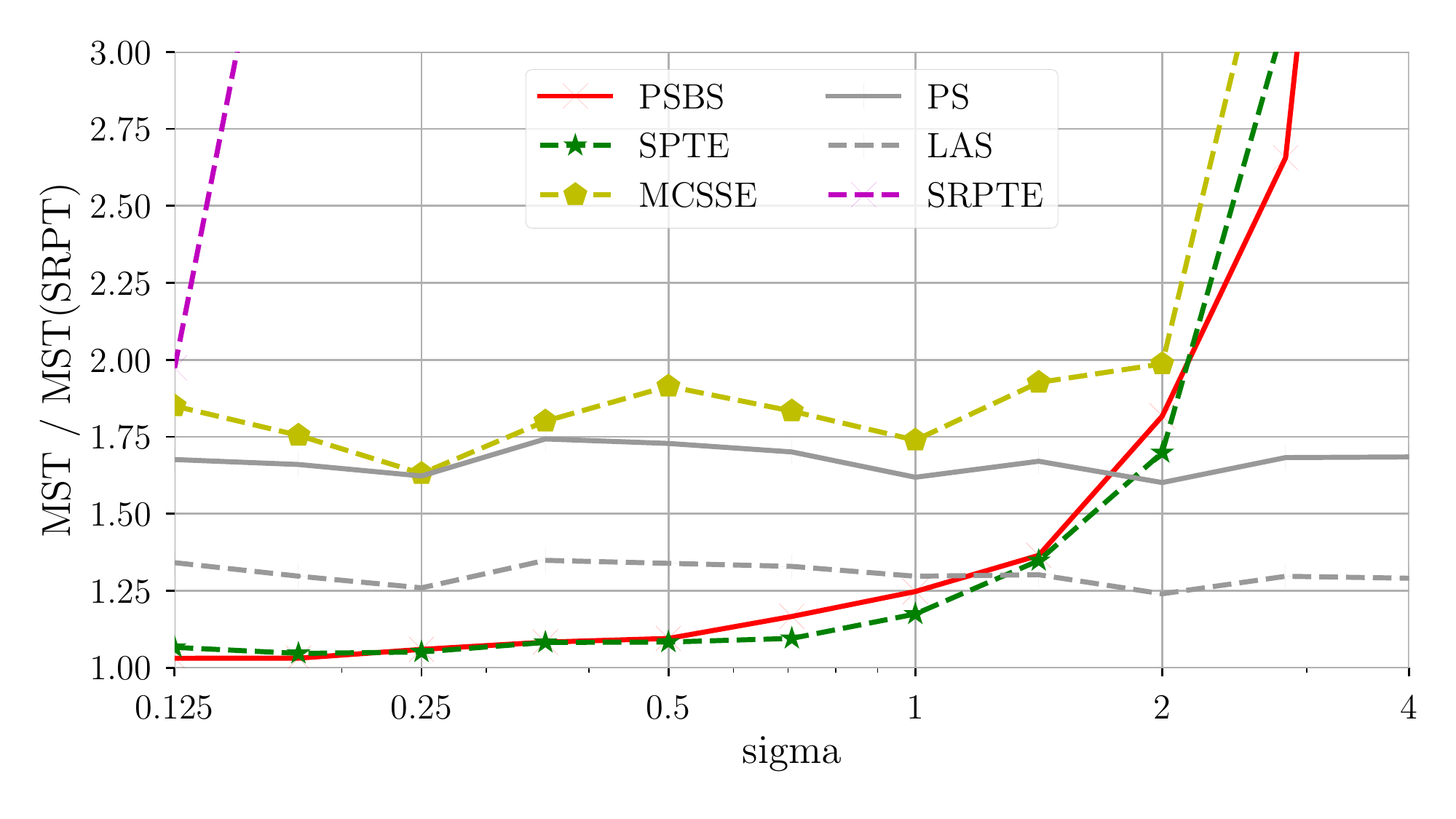}%
        \label{fig:sigma3}}
    \caption{Impact of error on heavy-tailed workloads, sorted by growing skew.}
    \label{fig:sigma}
\end{figure}

As we have seen in Figure~\ref{fig:3d}, the performance of size-based schedulers is significantly impacted by estimation errors only when the job size distribution is skewed. In Figure~\vref{fig:sigma}, we take three job size distributions sorted by growing skew, and we evaluate the performance of our algorithms. We see that, while SRPTE suffers the combination of job size skew and estimation errors, once again the performance figures of PSBS and SPTE are very close; interestingly, as skew grows, PSBS and SPTE's performance intersects with the one of PS at higher values: for shape=0.25 and 0.177, they are equivalent to PS for a sigma around 1.5; for an extremely skewed job size distribution having shape=0.125, this value moves to around 2. CSE is always outside of the plot for these values (see also Figure~\ref{fig:cse_3d}), while as long as estimation error is tolerable MCSSE performs similarly to PS.

\subsection{Fairness}

Fairness is a quite elusive concept in the context of scheduling~\cite{wierman2011fairness}: here, we continue to follow the approach of~\cite{dell2016psbs} and use slowdown (the ratio between a job's sojourn time and its size) as our main metric of fairness, according to the assumption that in a fair system a job's response time should be roughly proportinal to its size. In the following results, all parameters are fixed to their default values of Table~\ref{tab:params}.

\begin{figure}[!htbp]
    \centering
    \includegraphics[width=\columnwidth]{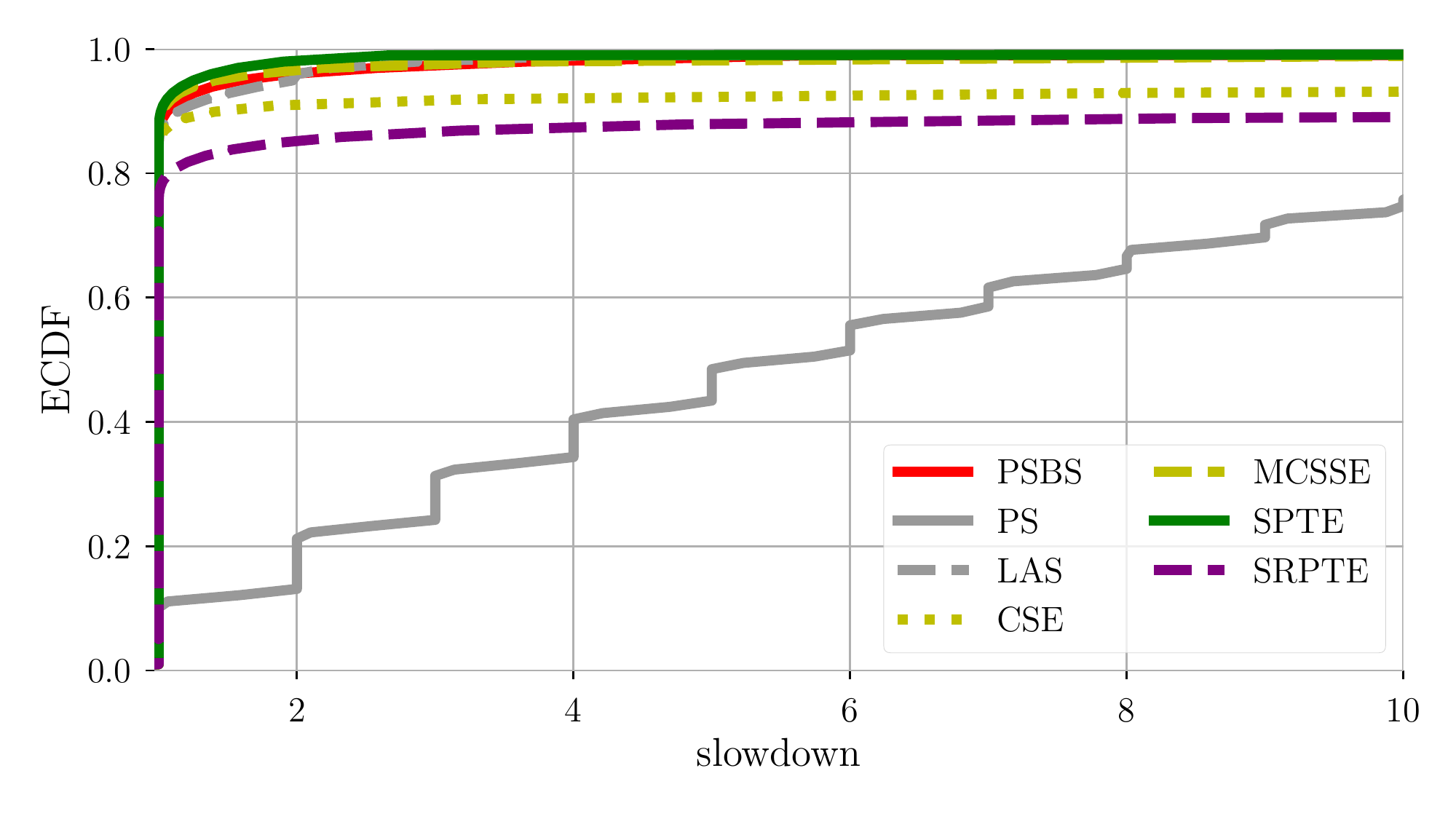}%
    \\
    \includegraphics[width=\columnwidth]{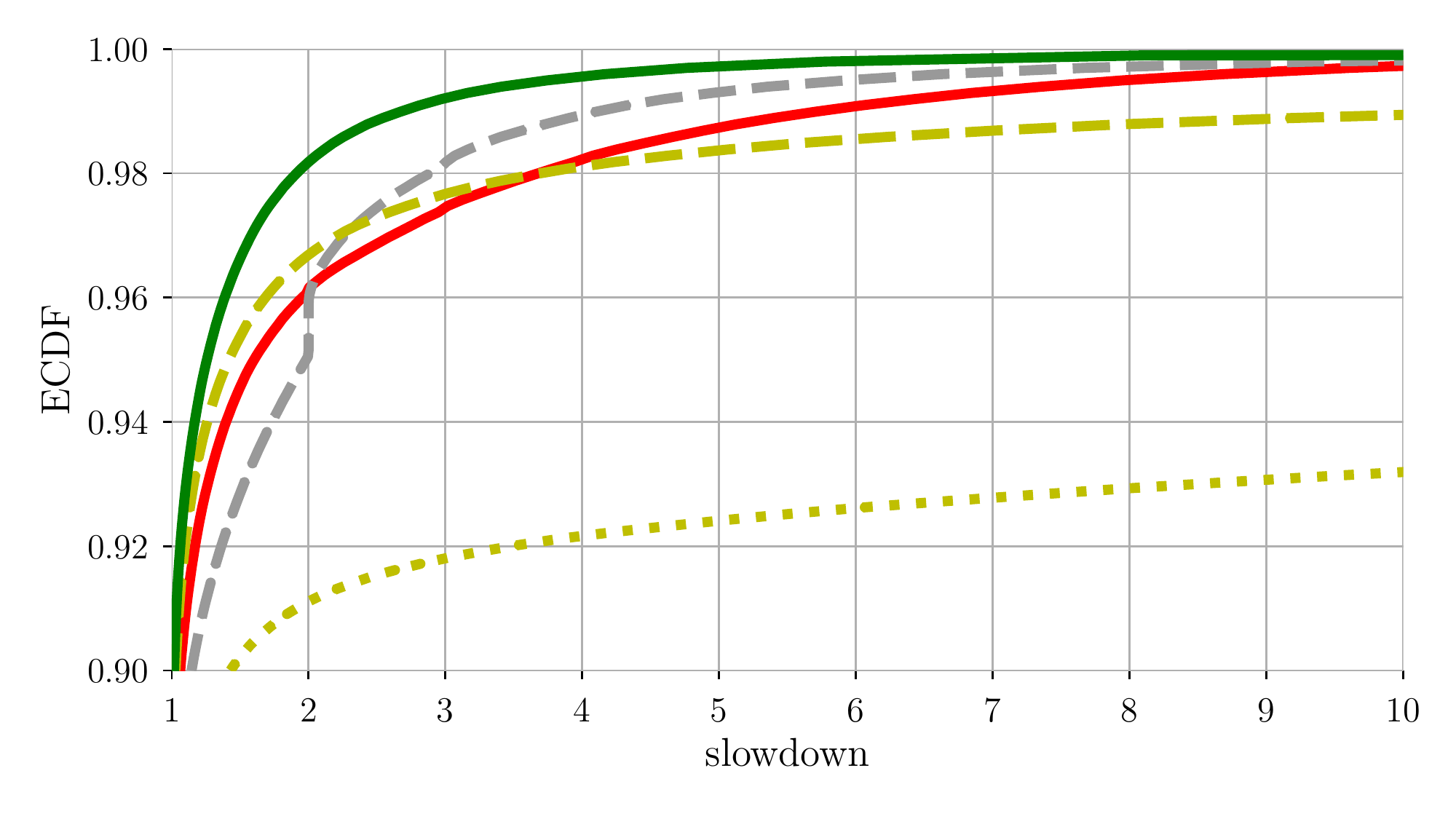}%
    \caption{Per-job slowdown: full CDF (top) and zoom on the 10\% more critical cases (bottom).}
    \label{fig:slowdown}
\end{figure}

Figure~\vref{fig:slowdown} shows the cumulative distribution function (CDF) of slowdown for all jobs observed in our experiments; the second figure zooms in on the 10\% worst values. While PS is often considered a standard for fairness, its slowdown distribution is quite wide: in a heavily loaded system jobs are heavily slowed down. Other algorithms perform better, with only a small fraction of jobs experiencing slowdowns worse than 2 or 3. It is perhaps surprising to observe that SPTE appears to outperform even PSBS in this case, with more than 98\% of jobs having a slowdown lower than 2: it appears that the policy of preempting large jobs even more aggressively pays off in this case, guaranteeing near-optimal slowdown values to even more jobs. It is interesting to see how LAS, in this case, perform very well obtaining results comparable to the ones of the best size-based algorithms.

\begin{figure}[!htbp]
    \centering
    \includegraphics[width=\columnwidth]{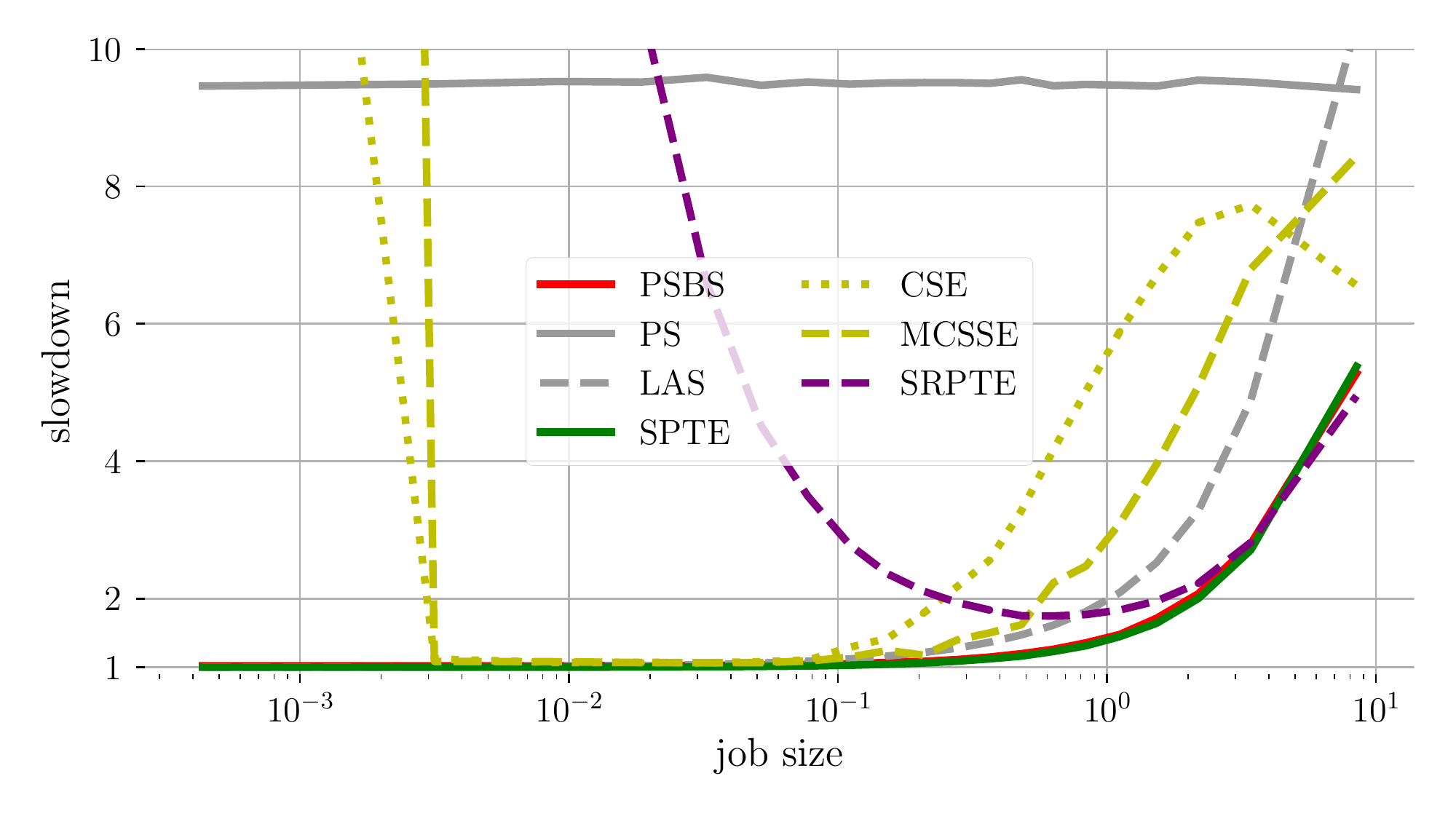}
    \caption{Mean conditional slowdown.}
    \label{fig:mcs}
\end{figure}

Another interesting way of defining fairness is \emph{mean conditional slowdown} (MCS), i.e., the average slowdown observed for jobs having a given size; a scheduling algorithm can be considered fair if MCS does not vary much between jobs having different sizes~\cite{wierman2007fairness}. In Figure~\vref{fig:mcs} we sort jobs, put them in 20 bins with similar size, and plot the average size versus the average slowdown for each bin. As \citet{wierman2007fairness} predicts, the MCS of PS remains constant for different job sizes; on the other hand, it is also noteworthy how PSBS and SPTE "dominate" PS by obtaining better results for each job size. It is interesting to see how the MCS of LAS is worse than the one of PS only for the very largest jobs, and that CSE and MCSSE perform badly for the very small jobs because they are likely to be somehow penalized by waiting for the completion of other small jobs that are still larger than them.

\subsection{Real Workloads}

\begin{figure}
    \centering
    \subfloat[FB09-0.]{%
        \includegraphics[width=\columnwidth]{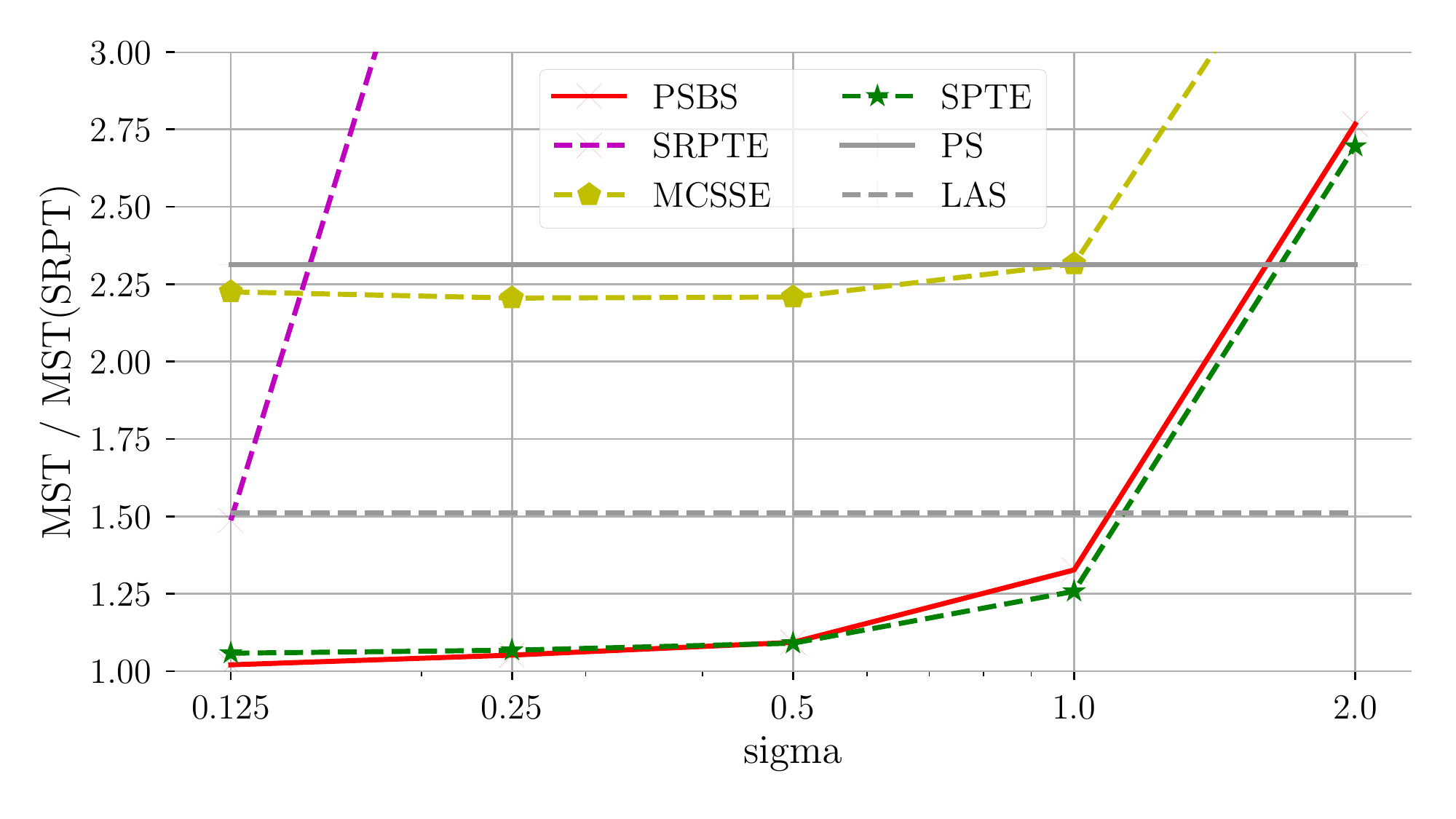}
        }
    \\
    \subfloat[FB09-1.]{%
        \includegraphics[width=\columnwidth]{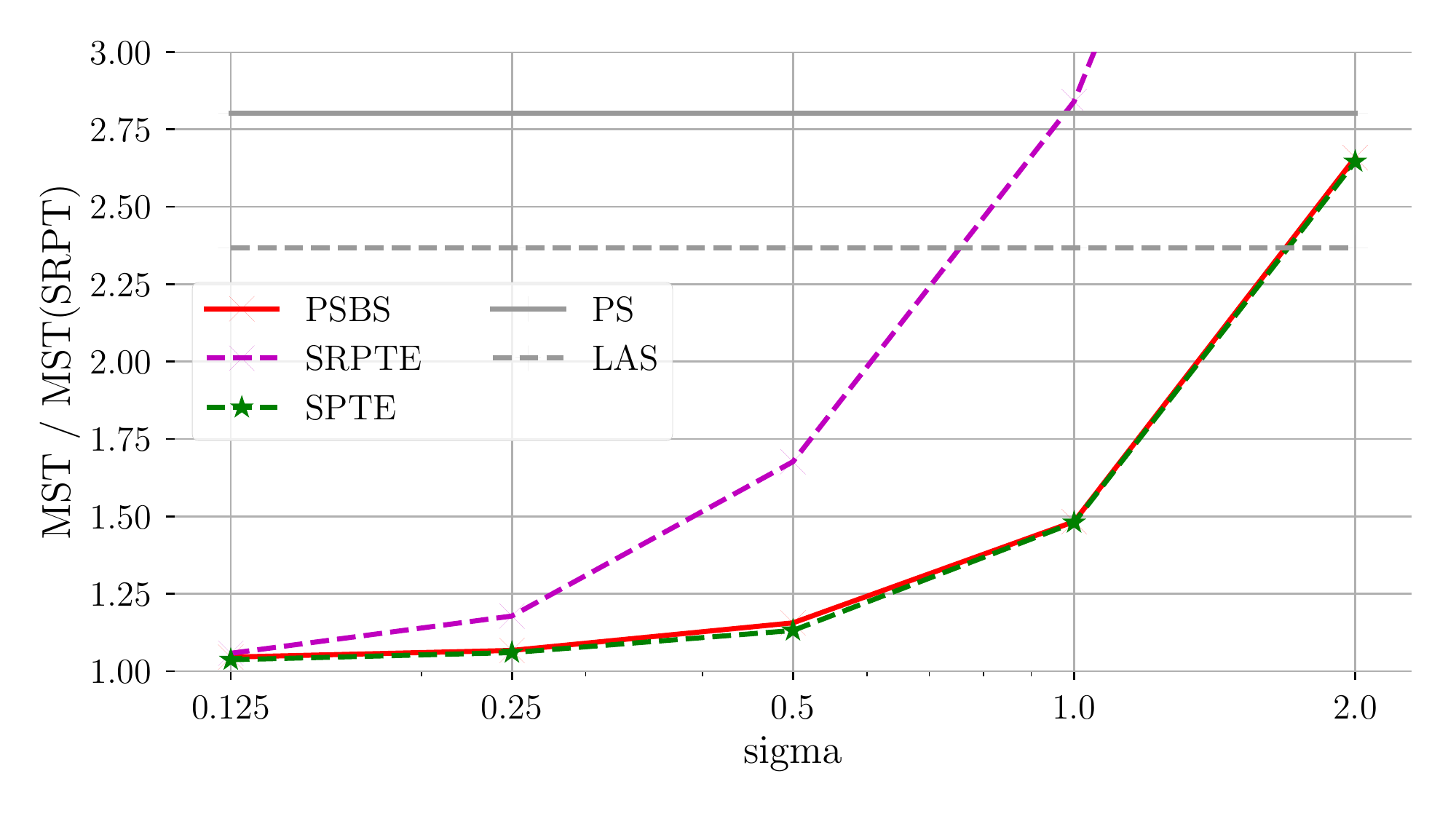}%
        }
    \\
    \subfloat[FB10.]{%
        \includegraphics[width=\columnwidth]{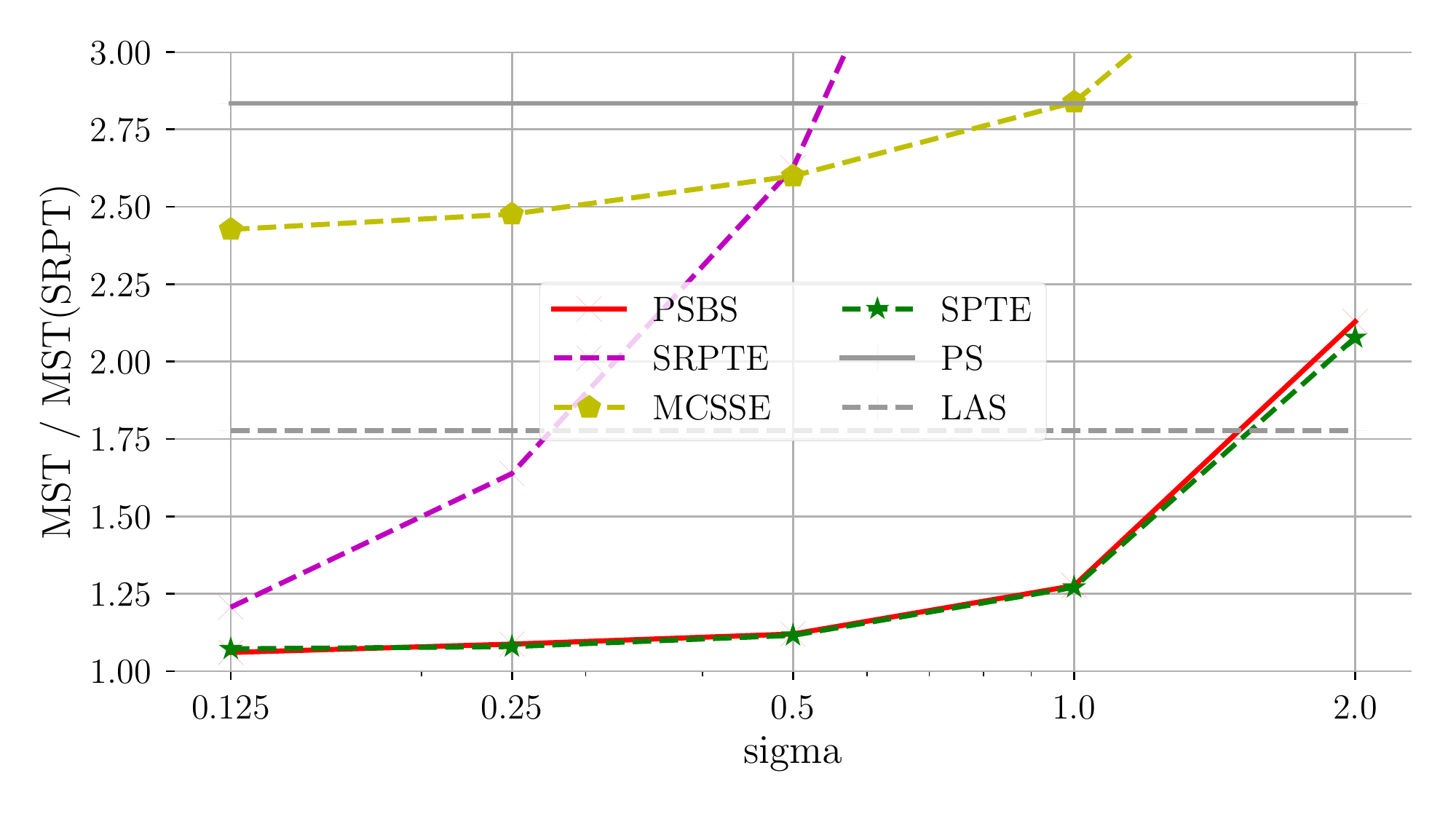}%
        }
    \caption{Evaluation on real workloads from Facebook.}
    \label{fig:FB}
\end{figure}

We finally cross-validate our findings by comparing our results on synthetic workloads to those that we obtain on three MapReduce workloads, made available through the SWIM tool~\cite{chen2012interactive}. The workloads have been parsed according using the procedure described in~\cite{dell2013simulator,dell2016psbs}. Confirming the findings of~\cite{dell2016psbs}, in Figure~\vref{fig:FB} we see that, for these skewed workloads, results are qualitatively similar to those for the synthetic Weibull distributions of Figure~\vref{fig:sigma}. In particular, we confirm that PSBS and SPT, in the presence of errors, behave once again in a similar fashion, obtaining an MST which is often close to optimal and preferable other policies in cases of high skew---unless errors are very large.

\section{Discussion}
\label{sec:discussion}

The experimental results show that, with respect to the metrics of performance (MST) and fairness (slowdown/MCS), SPTE performs better than MCSSE and similarly to PSBS. It is interesting to dig into these results, to provide an intuitive explanation to them.

First, the algorithms considered here generally perform acceptably (at least comparably to PS), while several other algorithms can do much worse, in particular for high skew and estimation error~\cite{dell2016psbs}: here, underestimating large jobs is not catastrophic because of either the ad-hoc modifications introduced by MCSSE and PSBS, or the fact that (unlike SRPT) priority simply never increases for any job in SPT.

Second, MCSSE may be seen as an approximation of SPTE---jobs are essentially binned in one of $r$ static priority classes, with the expectation that each priority class will be composed of jobs having similar size, and that smaller jobs will be scheduled first. As Figure~\ref{fig:shape} testifies, though, MCSSE performs comparatively worse in sweked workloads; when job size can vary by several orders of magnitude, priority classes become less and less homogeneous, and binning them into classes throws away useful information about job size; SPTE does not have this weakness.

Third, unlike SPTE, PSBS gives guarantees against starvation: as time passes, jobs increase their priorities and they will eventually be scheduled. One may wonder why the phenomenon of starvation does not appear in our results of Figures~\ref{fig:slowdown} and~\ref{fig:mcs}; this can be explained by the fact that in several cases, and in particular for skewed job size distributions, starvation for a size-based policy like SRPT is actually unlikely, and slowdown is generally lower than with a supposedly ``fair'' policy like PS~\cite{harchol1998case,Bansal:2001:ASS:384268.378792}. Our results imply that the same applies to SRPT's simpler relative, SPT.

\section{Conclusion}
\label{sec:conclusion}

Size-based scheduling algorithms could help in a several practical cases, but their real-world usage is limited by various factors: the impact of job size estimation errors, the complexity in implementing them together with a sound job size estimation framework, and the lack of strong analytic results on their performance. The simplicity of SPT may help for all these factors: besides the simplicity of implementation, the robustness to even large size estimation errors leads to making simpler estimation frameworks acceptable. In addition, it is possible that the simplicity of the algorithm itself may lead to easier and stronger theoretical results on its properties.

\balance

\bibliographystyle{IEEEtranN}
\bibliography{bibliography,references,zotero}

\end{document}